\documentclass[article,twocolumn,prb,showpacs,superscriptaddress]{revtex4-2}

\usepackage{graphicx}
\usepackage{amssymb}
\usepackage{bm}
\usepackage{float}
\usepackage{notes2bib}
\usepackage{pifont}
\usepackage{amsmath}
\usepackage{xfrac}
\usepackage{color}
\usepackage{titlesec}
\renewcommand{\figurename}{Figure} 
\makeatletter	        
\def\fnum@figure{\textbf{\figurename~\thefigure}}
\makeatother
\usepackage[normalem]{ulem}
\usepackage[utf8]{inputenc}
\usepackage[T1]{fontenc}
\usepackage{gensymb}
\usepackage[colorlinks=True,
                 linkcolor=black,
	        citecolor=blue,
	        urlcolor=blue]{hyperref}

\bibnotesetup{
note-name = ,
use-sort-key = false
}

\makeatletter	        
 \def\section{%
  \@startsection{section}{1}{\z@}{0.8cm plus1ex minus.2ex}{0.2cm}%
  {%
   \small\sffamily\bfseries\selectfont
   \raggedright
   \parindent\z@
  }%
 }%
  \def\subsection{%
  \@startsection{subsection}{2}{\z@}{0.8cm plus1ex minus.2ex}{0.2cm}%
  {%
   \small\sffamily\bfseries
   \raggedright
   \parindent\z@
  }%
 }%
\makeatother

\newcommand{\comment}{\textcolor{black}}

\usepackage{color}

\newcommand{\MOE}{MOE Key Laboratory for Nonequilibrium Synthesis and Modulation of Condensed Matter, Shaanxi Province Key Laboratory of Advanced Materials and Mesoscopic Physics, School of Physics, Xi’an Jiaotong University, Xi’an,710049, China}

\makeatletter
\g@addto@macro\bfseries{\boldmath}
\makeatother

\begin{document}
\title{Inducing Berry Curvature Dipole in \comment{Multilayer} Graphene through Inhomogeneous Interlayer Sliding}
\author{Jie Pan}
\author{Huanhuan Wang}
\author{Lin Zou}
\author{Haibo Xie}
\author{Yi Ding}
\author{Yuze Zhang}
\author{Aiping Fang}
\author{Zhe Wang}
\email{zhe.wang@xjtu.edu.cn}
\affiliation{\MOE}

\begin{abstract}

Breaking lattice symmetry is crucial for generating a nonzero Berry curvature. While manipulating twisting angles between adjacent layers has successfully broken lattice symmetry through strain field and generated nonzero Berry curvature, interlayer sliding in principle offers a promising alternative route. However, realizing uniform interlayer sliding faces experimental challenges due to its energetic instability. In this work, we introduce an experimentally feasible method, using a corrugated substrate to induce an \comment{inhomogeneous but energetically more stable} interlayer sliding in \comment{multilayer} graphene. Our simulations demonstrate that \comment{inhomogeneous interlayer sliding} produces a sizable Berry curvature dipole, which can be further tuned by varying the interlayer sliding distances and potential differences. The resulting Berry curvature dipole magnitude is remarkably up to 100 times greater than the maximum displacement involved in the inhomogeneous sliding. Our results highlight \comment{inhomogeneous} interlayer sliding as a viable and effective method \comment{to induce a significant Berry curvature dipole in graphene systems and propose the experimentally feasible way to realize it.}

\end{abstract}

\maketitle

\section*{I.	Introduction}

The Hall effects\cite{HE} — including anomalous Hall effect(AHE) \cite{AHERMP}, quantum Hall effect (QHE) \cite{QHE} and fractional quantum Hall effect (FQHE) \cite{FQHE} — require broken time-reversal symmetry, which is typically achieved by applying magnetic field or inducing magnetization. Investigations into the mechanisms underlying AHE and QHE have contributed significantly to the development of topological physics, revealing that these effects are closely related to Berry curvature \cite{Berry,Niu2010,Berry2}. Recent study\cite{Fu2015} has shown that nonzero Berry curvature dipole, defining as the first moment of the Berry curvature in momentum ($k$) space\cite{Fu2015,Du2018,Battilomo2019}, can induce nonlinear Hall effect in systems that preserve time-reversal symmetry.  Unlike conventional Hall effect, this nonlinear Hall effect is a second-order response, where transverse voltage $V_{xy} \propto I^2$, with $I$ being the applied alternating current. Importantly, a nonzero Berry curvature dipole requires further reduced spatial symmetry; for instance, in two-dimensional (2D) systems, at most mirror symmetry can be maintained\cite{Fu2015,Battilomo2019}. Significant efforts have been directed toward breaking certain spatial symmetries to achieve a nonzero Berry curvature dipole. One approach is to identify intrinsically low-symmetry materials, such as bilayer and few-layer WTe$_{2}$\cite{Du2018,Ma2019,Mak2019}. Another strategy involves strain engineering to break the three-fold rotational symmetry, characteristic of graphene and many 2D transition metal dichalcogenides (TMDCs). This can be achieved by applying strain in graphene \cite{Battilomo2019,Ho2021,Guinea2021,Sinha2022,Law2022,Huang2023PRL} and TMDCs such as MoS$_2$\cite{Son2019PRL}, WSe$_2$ \cite{Hu2022, Huang2022NSR}, etc.

The spatial symmetry of 2D materials is determined by the atomic lattice registry, which can be tuned by two crucial parameters,  the twisting angle and the sliding vector. In bilayer graphene system, manipulating twisting angle has become a popular and effective experimental method \cite{Cao2018mott,Cao2018sc} to modulate the electronic properties and topological phases. Various exotic phenomenon emerges, including Mott insulator-metal transition\cite{Cao2018mott}, unconventional superconductivity \cite{Cao2018sc}, ferromagnetism \cite{sharpe2019}, topological phases\cite{Choi2021,Saito2021}, etc. It is important to note that rigid twisting alone does not break the three-fold symmetry in graphene or TMDCs; the strain field induced by twisting is the primary factor responsible for symmetry breaking \cite{Guinea2021, Hu2022,Law2022}. Similarly, the sliding vector plays a critical role in determining spatial symmetry. While studies on interlayer sliding have primarily focused on theoretical calculations of band structures \cite{Son2006,Park2015,Lee2015,Nam2021}, we recently showed that uniform interlayer sliding can effectively break the three-fold rotational symmetry of carbon lattices, thereby inducing a significant Berry curvature dipole \cite{pan2024}. \comment{On the experimental side, uniform interlayer sliding has been realized in ferroelectric materials, such as manually parallelly stacked h-BN \cite{Stern2021,Pablo2021}, TMDCs \cite{ferro1,ferro2,ferro3}. But for nonpolar multilayer graphene system, realizing uniform interlayer sliding remains challenging\cite{csliding1,csliding2}.}



\begin{figure*}[t]
\includegraphics[width =0.8\linewidth]{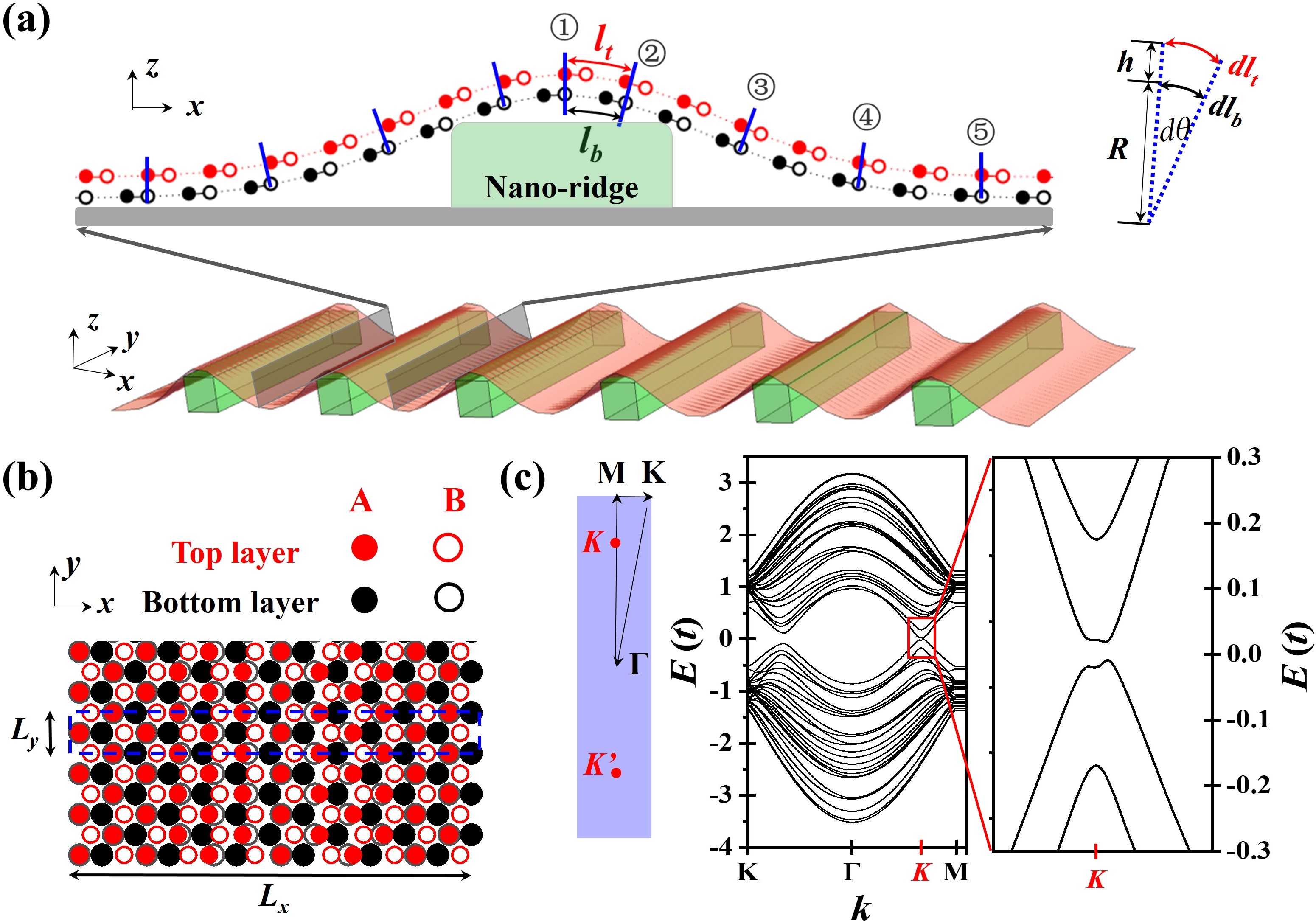}
\caption{\textbf{Simulated bilayer graphene with inhomogeneous interlayer sliding.} (a) Schematic illustration of generating \comment{inhomogeneous interlayer sliding in bilayer graphene by placing it on top of an array of nano-ridges.}  (b) Carbon lattice structure with inhomogeneous interlayer sliding, where the blue dashed line outlines the unit cell used for simulation. (c) Brillouin zone and band structure of bilayer graphene with inhomogeneous interlayer sliding. The red markers denote the $K$ and $K'$ valleys of graphene, which are folded into the first Brillouin zone.}
\label{cartoon}
\end{figure*}

\comment{Recent} experimental observation of local interlayer sliding \comment{in bent graphene is inspiring.} \comment{When multilayer graphene is} placed on h-BN steps\cite{Han2020}, \comment{inhomogeneous sliding is induced and the largest} sliding distance can be $\sim2a$ (with $a$=0.142 nm being carbon-carbon bond length). \comment{However, the induced inhomogeneous interlayer sliding breaks translational symmetry along the sliding direction (the $x$ direction in our case). To restore the translational symmetry, such inhomogeneous sliding must be periodic, enabling the definition of $k_x$.} \comment{In this work, we} propose to induce a nonzero Berry curvature dipole in \comment{multilayer} graphene by placing it on corrugated substrate, which is made of an array of nano-ridges as shown in Fig. \ref{cartoon}. \comment{This nano-ridge array creates periodic} inhomogeneous interlayer sliding in \comment{multilayer} graphene, \comment{which in principle can be controlled by the geometry of nano-ridge}. Our simulations indicate that this experimentally feasible method can produce a significant and tunable Berry curvature dipole.

In what follows, we introduce the inhomogeneous interlayer sliding model in Section II. The simulation results, including the distributions of Berry curvature and Berry curvature dipole density are discussed is Section III. We investigate how the Berry curvature dipole varies with Fermi energy and explore the effects of interlayer potential differences, interlayer sliding distances, \comment{layer numbers} and system length scales. A brief summary is included in Section IV.

\section*{II.	Simulation Methods}

As Bernal (AB) stacking is the most stable configuration for \comment{multilayer} graphene, inducing a uniform interlayer sliding will result in an increase in energy. To address this challenge, we propose to put \comment{multilayer} graphene on corrugated substrates, for instance \comment{a nano-ridge. This setup allows for the natural induction of inhomogeneous interlayer sliding, as illustrated in Fig. \ref{cartoon}(a). In this configuration, solid (open) circles represent the A(B) carbon lattices, with red (black) colors denoting the top (bottom) layers. The blue lines indicate the local perpendicular (out-of-plane) direction. Due to the much stronger intralayer $\sigma$ bonds compared to the interlayer $\pi$ bonds, we neglect variations in $\sigma$ bond lengths caused by bending bilayer graphene. This deformation results in a difference in path lengths, $l_t>l_b$, between the top and bottom layers, thereby inducing interlayer sliding and altering the local atomic registry. Such interlayer sliding caused by bending has been experimentally observed in multilayer graphene \cite{Han2020}, and a similar mechanism has been theoretically explored in $h$-BN \cite{bend}. Starting with AB stacking at the nano-ridge center (region \ding{172}), the stacking configuration deviates slightly in region \ding{173}. The interlayer sliding reaches its maximum in region \ding{174} and gradually decreases as the distance from the nano-ridge increases, eventually returning to AB stacking in region \ding{176}. A similar pattern of interlayer sliding occurs on the opposite side of the nano-ridge.}

\comment{The analytical expression for the interlayer sliding distance can be derived by calculating $l_t-l_b$.  Considering the infinitesimal arcs, as illustrated in the right panel of Fig. \ref{cartoon}(a), the infinitesimal sliding distance is given by: $ds=dl_t-dl_b=hd\theta$, where $h$ is the layer separation, approximated as $\sim 2.4a$ for bilayer graphene \cite{Park2015}. Integrating over the deformation profile, the total interlayer sliding induced by an arbitrary deformation can be quantified as:}
\begin{equation}
    s = l_t-l_b = h \arctan \left(-\frac{{\rm d}z}{{\rm d}x}\right),
\label{sliding1}
\end{equation}
\comment{where $\frac{{\rm d}z}{{\rm d}x}$ denotes the slope of deformed structure. Experimentally, bending tetralayer graphene($h=7.2a$) by 12\degree, has been shown via TEM to induce interlayer sliding of approximately $a$ to $2a$\cite{Han2020}, which aligns well with the predicted value of  $1.5a$ based on Eq. \ref{sliding1}, providing further validation of its accuracy. Given that the slope $\frac{{\rm d}z}{{\rm d}x}$ varies with $x$, bending multilayer graphene is expected to induce inhomogeneous interlayer sliding.} For simplicity, we assume the height profile of the \comment{multilayer} graphene follows a Gaussian function,
\begin{equation}
    z(x) = h_0 \exp \left( {\frac{-x^2}{2\sigma^2}} \right),
\label{shape}
\end{equation}
where $h_0$ and $\sigma$ is determined by the geometry of the structure, namely the height and width of nano-ridge. Substituting Eq. \ref{shape} into Eq. \ref{sliding1}, we can show the maximum sliding happens at $x = \sigma$, with maximum sliding distance to be,
\begin{equation}
   s_m = h \arctan \left(\frac{h_0}{\sigma \sqrt{e}}\right).
\label{sliding2}
\end{equation}
We use the maximum shift $s_m$ to characterize the overall sliding magnitude of the system.

\comment{To achieve periodic structures in $x$ direction of the multilayer graphene, we propose to use the arrays of nano-ridges to induce the periodic interlayer sliding, as illustrated in Fig. \ref{cartoon}(a).} For calculations we focus on a representative unit cell, as illustrated by Fig. \ref{cartoon}(b), where we take bilayer graphene with periodic interlayer sliding for demonstration. $L_{x(y)}$ denotes the periodicity in $x(y)$ directions. We fix $L_x=18a$ and investigate the size-scaling behavior later in this study. It is important to note that the supercell in Fig. \ref{cartoon}(b) represents the interlayer sliding induced by deformation in three-dimensional space, which affects only the interlayer coupling. The carbon lattice within each layer remains unchanged, preserving the intralayer hopping parameters.

To model the interlayer hopping energy variation under interlayer sliding, we adopt the following parameterization, which is widely used as in Refs \cite{Ando2001,Uryu2004,Trambly2010,Koshino2012},
\begin{equation}
t\left( {\boldsymbol{\delta },z} \right) = {V_\pi } e^  { - \left( {d - a} \right)/{r_0}} {\sin ^2}\alpha  + {V_\sigma } e^ { - \left( {d - h} \right)/{r_0}} {\cos ^2}\alpha
\label{hopping}
\end{equation}
where $V_\pi = - t$ denotes the intralayer hopping component from $\pi$ bond, $V_\sigma = 0.18t$ representing the interlayer hopping component from $\sigma$ bond, $r_0 = 0.0453$ nm is the decaying length. $d$ is the separation between two atoms, defining as  $d =\sqrt{\left| \boldsymbol{\delta} \right|^2+z^2}$ with $\boldsymbol{\delta}$ and $z$ to be the intralayer and interlayer displacement component respectively, and $\alpha$ is the angle between displacement vector $(\boldsymbol{\delta}, z)$ and perpendicular $\boldsymbol{z}$ axis.

By substituting Eq. \ref{hopping} to the tight binding model, we can obtain the Hamiltonian
\begin{equation}
H = \sum_{\{i,j\}} t\left(\bm {r_{a,i}-r_{b,j}} \right)\left(c_i^\dagger c_j +c.c.\right) + \sum_{l,i} V_l c_{l,i}^\dagger c_{l,i} .
\label{ham}
\end{equation}
where $c^\dagger$ and $c$ denotes creation and annihilation operators respectively. The first summation covers all the neighbor hopping with a intralayer cut-off distance $\delta=1.5a$. \comment{The second summation introduces a layer-dependent potential $V_l$, where the subscript $l$ denotes the layer index. For an $n$-layer graphene system, we set $V_l=V\cdot(2l-n-1)$. Specifically, in the case of bilayer graphene, this results in $V_{1}=-V$ for the first layer and $V_{2}=V$ for the second layer.} With a periodic boundary condition, the Hamiltonian Eq. \ref{ham} can be used for bandstructure and wavefunction simulations.

\section*{III.	Results and Discussions}

The first Brillouin zone of the supercell is shown in the left panel of Fig. \ref{cartoon}(c), where K, M and $\Gamma$ represents the high-symmetric points. Since our focus is on the low-energy region, it is important to identify the $K$ and $K'$ valleys of graphene. Due to zone folding, these $K$ and $K'$ valleys are folded into the first Brillouin of the supercell as indicated by the red point. 

The bandstructure is obtained by diagonalizing Hamiltonian in Eq. \ref{ham}. For the numerical calculations, we set $s_m = 0.2a$ and $V=0.02t$ for top and bottom layers. The simulated bandstructure is shown in Fig. \ref{cartoon}(c), where the wave vector $k$ varies along the path from K to $\Gamma$, to $K$ to M and finally back to K. The right panel presents a zoomed-in view of the bandstructure, clearly showing a band gap opening at the $K$ valley. \comment{In the following, we focus initially on the bilayer graphene case to investigate the Berry curvature dipole as a function of Fermi energy, sliding distances, and interlayer potential differences. Subsequently, in Subsection E, we extend our analysis to trilayer and tetralayer graphene. Finally, in Subsection F, we explore the scaling behavior with system size.}

\subsection{Berry curvature \& dipole density}

\begin{figure*} [t]
\includegraphics[width =0.9\linewidth]{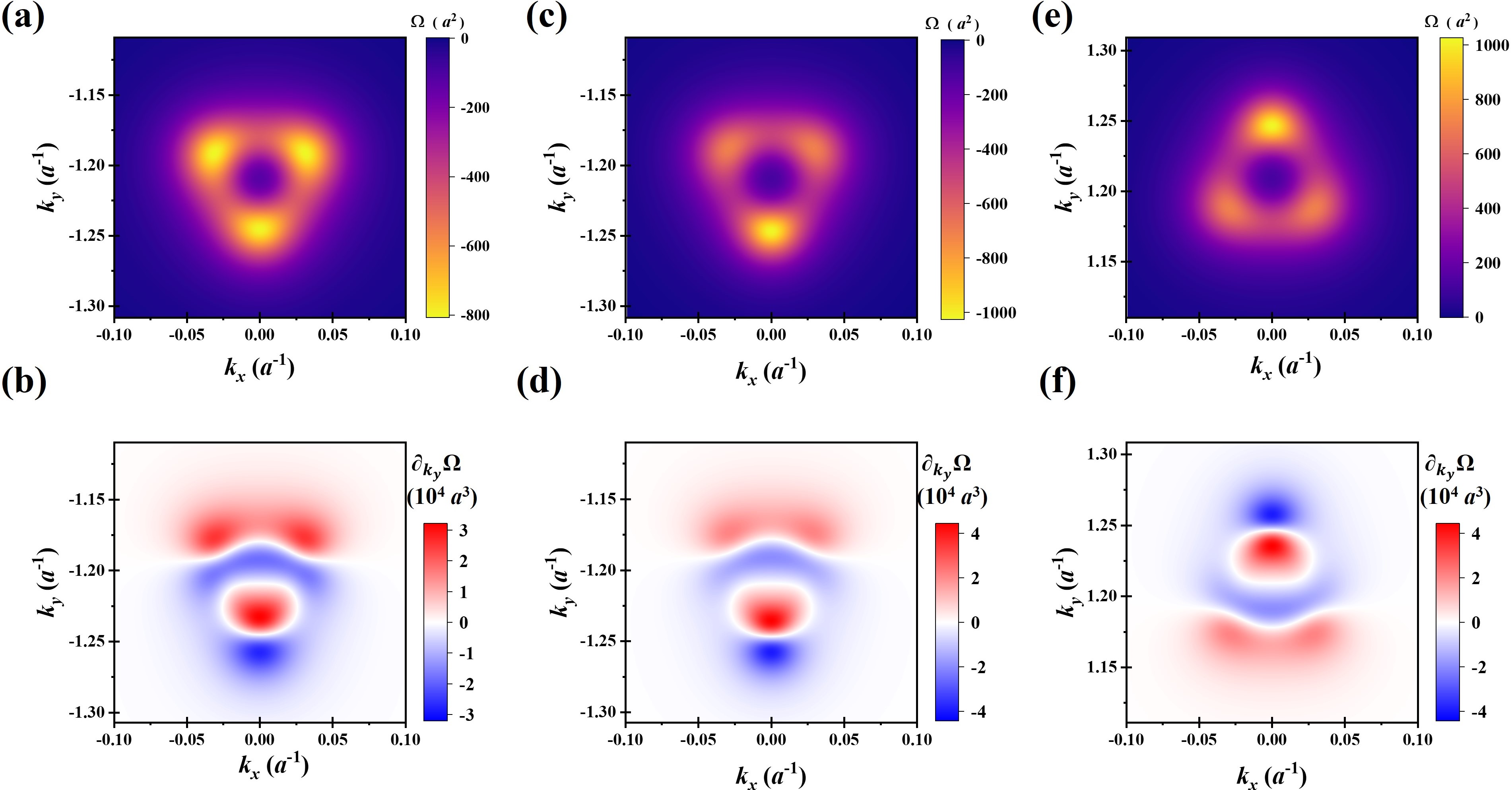}
\caption{\textbf{Simulated Berry curvature and dipole density.} The Berry curvature and dipole density in the $k$ space are illustrated for three cases:(a) and (b) show the Bernal stacking ($s_m=0$) case near the $K'$ valley, (c) and (d) show the $s_m=0.2a$ case near $K'$ valley, and (e) and (f) show the $s_m=0.2a$ case near $K$ valley. The potential difference is fixed at $V = 0.02t$ for all cases.}
\label{bc}
\end{figure*}

The Berry curvature is calculated based on the following definition \cite{Berry,Niu2010},
\begin{equation}
{\Omega_c} = i\frac{{\sum_{j \ne c} {\left[ {\left\langle {{\phi _j}} \right|\partial H/\partial {k_x}\left| {{\phi _c}} \right\rangle \left\langle {{\phi _c}} \right|\partial H/\partial {k_y}\left| {{\phi _j}} \right\rangle  - c.c.} \right]} }}{{{{\left( {{\varepsilon _c} - {\varepsilon _j}} \right)}^2}}},
\label{equ_bc}
\end{equation}
where the index $c$ refers to the index of first conduction band, $\phi_j$ and $\varepsilon_j$ denotes the $j$th eigenvectors and eigenvalues, respectively. We set the interlayer potential difference $V = 0.02t$, around 60 meV which is attainable by applying a displacement field using dual gates \cite{Zhang2009, Pablo2010}. We first evaluate the Berry curvature $\Omega$ near $K'$ valley without interlayer sliding, $s_m = 0$, which is shown in Fig. \ref{bc}(a). For Bernal-stacked bilayer graphene, there are three nonzero Berry curvature pockets, which are related to the trigonal warping effect and Lifshitz transition of bilayer graphene\cite{trigonal}. Specifically, these features are related to the interlayer hopping between non-dimer sites (between carbon atom B on top layer and carbon atom A on bottom layer as shown in Fig. \ref{cartoon}(b)). More importantly, three-fold rotational symmetry is exhibited. Our numerical tight-binding results are the same as the continuum model as discussed in Ref. [\onlinecite{pan2024}].

When inhomogeneous interlayer sliding is induced, the three-fold rotational symmetry is expected to be broken. We simulated the Berry curvature distribution for the case with $s_m=0.2a$. Near the $K'$ point, we observed that the absolute magnitude of upper two nonzero Berry curvature pockets is smaller than that of the lower one, clearly exhibiting the broken three fold rotational symmetry, as illustrated in Fig. \ref{bc}(b). Additionally, at $K$ valley, the Hamiltonian obeys $H_{K}(k_x,k_y) = H_{K'}(k_x,-k_y)$, leading to the antisymmetric Berry curvature distribution with $\Omega_{K'}(k_x,k_y) = - \Omega_{K}(k_x,-k_y)$, as evidenced by Figs. \ref{bc}(c) and \ref{bc}(e).

The Berry curvature dipole density can be evaluated from the Berry curvature distribution by taking derivative in $k$ space, i.e, \comment {$\partial_{k_{x,y}} \Omega = \partial \Omega / \partial k_{x(y)}$} \cite{Battilomo2019}. The results for \comment{$\partial_{k_y} \Omega$} are summarized in Figs. \ref{bc}(b) for $s_m = 0$,(d) $s_m = 0.2a$ near $K'$ and (f) $s_m = 0.2a$ near $K$ valleys. It is important to notice in all cases shown in Fig. \ref{bc}, the Berry curvature density and $y$ component dipole density \comment{$\partial_{k_y} \Omega$} exhibit the mirror symmetry with respect to $k_x=0$, indicating that \comment{$\partial_{k_y} \Omega$} is an even function of $k_x$. This mirror symmetry arises because, within the same valley, the time-reversal symmetry of the Hamiltonian is preserved in the $x$ direction: $H(-k_x,k_y)=H^*(k_x,k_y)$. Substituting this relation into Eq. \ref{equ_bc} results in the observed mirror symmetry with respect to $k_x=0$. Consequently, \comment{$\partial_{k_x} \Omega$} is an odd function in $k$ space obeying $\partial_{k_x} \Omega(-k_x,k_y)=-\partial_{k_x} \Omega(k_x,k_y)$. Therefore, when integrated \comment{$\partial_{k_x} \Omega$} over occupied states, the Berry curvature dipole $D_x$ becomes zero. This point will be verified in the next subsection.

\subsection{Berry curvature dipole v.s. Fermi energy}

\begin{figure} [t]
\includegraphics[width =1\linewidth]{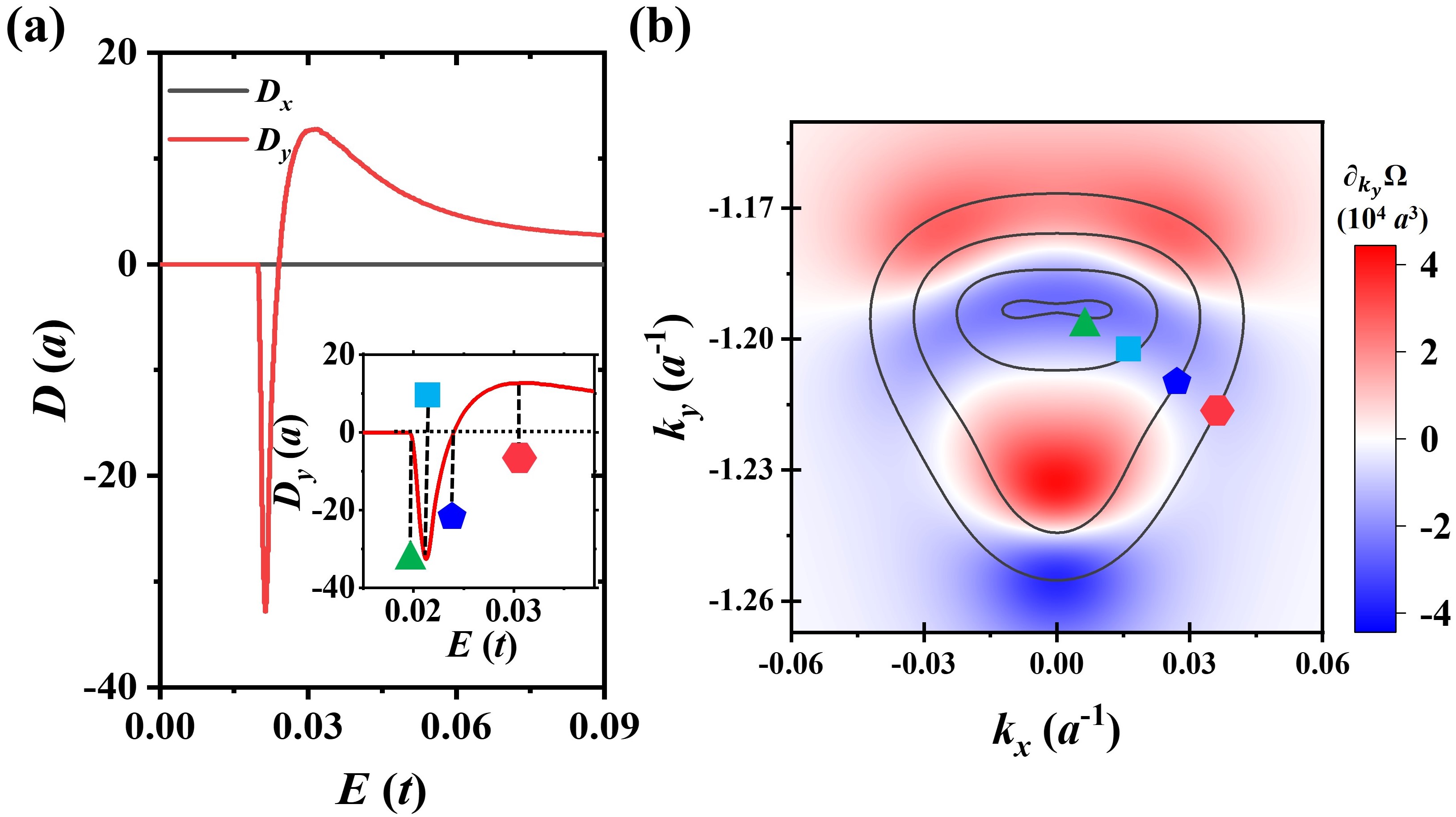}
\caption{\textbf{Simulated Berry curvature dipole as a function of Fermi energy.} (a) Berry curvature dipole $D_x$ and $D_y$ as functions of Fermi energy. Inset is the enlarged view of $D_y$ with four characteristic energies marked by distinct symbols. (b) Berry curvature dipole density \comment{$\partial_{k_y} \Omega$} is shown alongside four iso-energetic lines in $k$ space, corresponding to the same four symbols from the inset in (a). In the simulation, we set $V=0.02t$ and $s_m=0.3a$.}
\label{bcd}
\end{figure}

The Berry curvature dipole $D_y$ is defined as the integral of Berry curvature dipole density \comment{$\partial_{k_y} \Omega$} over occupied states \cite{Fu2015,Battilomo2019} as follows,
\begin{equation}
  {D_{x(y)}} = \int_k {f\left( \partial_{k_{x(y)}}\Omega \right)} = 2\int_k {f\left( \partial_{k_{x(y)}}\Omega_{K'} \right)}
\end{equation}
where $f$ is the Fermi-Dirac distribution, approximated as a step function in the zero-temperature limit; the coefficient 2 comes from the valley degeneracy. The evaluated Berry curvature dipole is plotted as a function of Fermi energy in Fig. \ref{bcd}. $D_x$ is zero, as the Berry curvature density \comment{$\partial_{k_x} \Omega$} is an odd function, leading to cancellation after integration. In contrast, the Berry curvature dipole $D_y$ is nonzero as shown by the red curve. The inset shows the enlarged view of $D_y$ curve, highlighting four critical energy points with distinct symbols. 

The sign-switching behavior can be understood with reference to Fig. \ref{bcd}(b), where the Berry curvature density and the iso-energetic lines are plotted together. As Fermi energy increases, we first encounter regions with negative Berry curvature dipole density, indicated by the iso-energetic curve marked by triangle symbol. This accounts for the initial appearance of a negative Berry curvature dipole. As the Fermi energy further increases, additional regions with negative Berry curvature dipole density contributes to the integral until reach a critical point marked by the square symbol. At this point, Berry curvature dipole reaches a maximum absolute magnitude. Further increases in Fermi energy introduces more regions with positive Berry curvature dipole density, leading to a transition to positive value after passing the energy marked by the pentagon symbol. Maximum Berry curvature is reached at energy of hexagonal symbol. After that, the negative Berry curvature dipole density in the lower part would result a decay of the Berry curvature dipole.

The above analysis clearly demonstrate that Berry curvature dipole is closely related to the Lifshitz transition in bilayer graphene\cite{trigonal}. It also indicates that by measuring the Berry curvature dipole (for instance through the nonlinear Hall effect), one can detect these complex topological phases. Similar discussions can also be found in twisted bilayer WSe$_2$ \cite{Hu2022} as well as bilayer graphene with uniform interlayer sliding \cite{pan2024}.

\subsection{Berry curvature dipole v.s. Sliding distance}

\begin{figure} [t]
\includegraphics[width =0.9\linewidth]{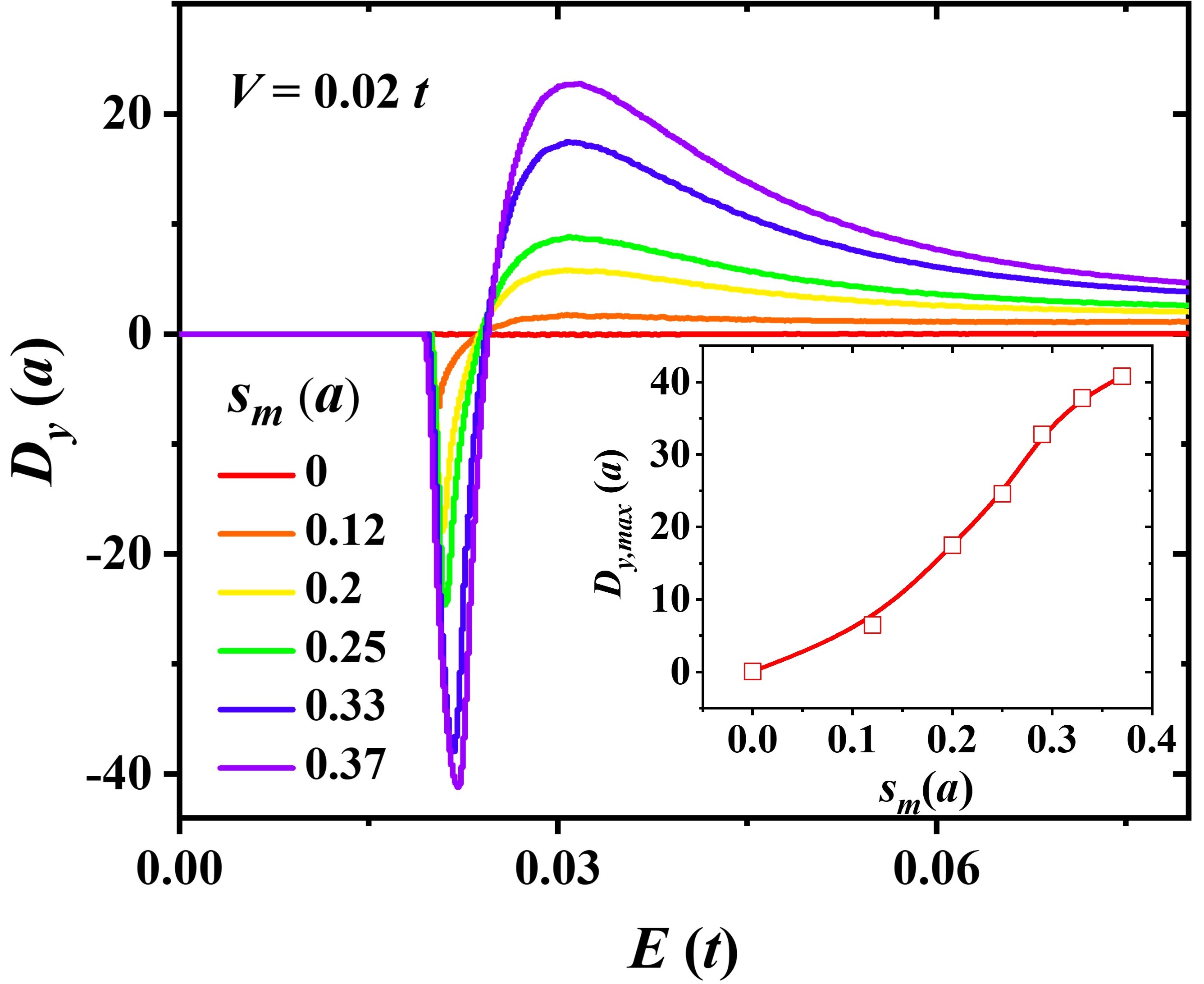}
\caption{\textbf{Simulated Berry curvature dipole under different sliding distances.} Berry curvature dipole is plotted as a function of Fermi energy for various sliding distances $s_m$, ranging from 0 (Bernal stacking) to $0.37a$. The inset summarizes the maximum magnitude of $|D_y|$ as a function of sliding distance $s_m$, with the red curve serving as a guide to the eye. The potential difference is fixed at $V=0.02t$.}
\label{distance}
\end{figure}

As discussed in subsection A, inducing interlayer sliding is crucial for breaking three-fold rotational symmetry to produce a nonzero Berry curvature dipole. In this subsection, we calculate the Berry curvature dipole with different sliding distances, varying from $s_m = 0$ to $s_m = 0.37a$. For each case, the interlayer potential difference is fixed as $V=0.02 t$, the Berry curvature dipole is plotted as a function of Fermi energy as illustrated in Fig. \ref{distance}. In the case of $s_m = 0$, the Berry curvature dipole is zero due to the preservation of three-fold rotational symmetry. For all other cases with interlayer sliding,  a similar behavior is observed: the Berry curvature dipole initially exhibits a negative value, gradually transitioning to a positive value and finally decays to zero, consistent with the analysis presented in Fig. \ref{bcd}.

To quantify the differences in magnitude, we use the maximum value of $|D_y|$, denote as $D_{y,{\rm max}}$, to represent the overall magnitude of the Berry curvature dipole. The results are summarized in the inset of Fig. \ref{distance}. It reveals a clear monotonic relationship with a slope of approximately 100. This indicates that a small sliding distance of $0.4a$ ($\sim$ 0.06 nm) can result in a significant Berry curvature dipole of $40a$ ($\sim$ 6 nm). These findings underscore that inducing inhomogeneous interlayer sliding is a highly effective method for generating a nonzero Berry curvature dipole in bilayer graphene.

\subsection{Berry curvature dipole v.s. Potential difference}

\begin{figure} [t]
\includegraphics[width =0.9\linewidth]{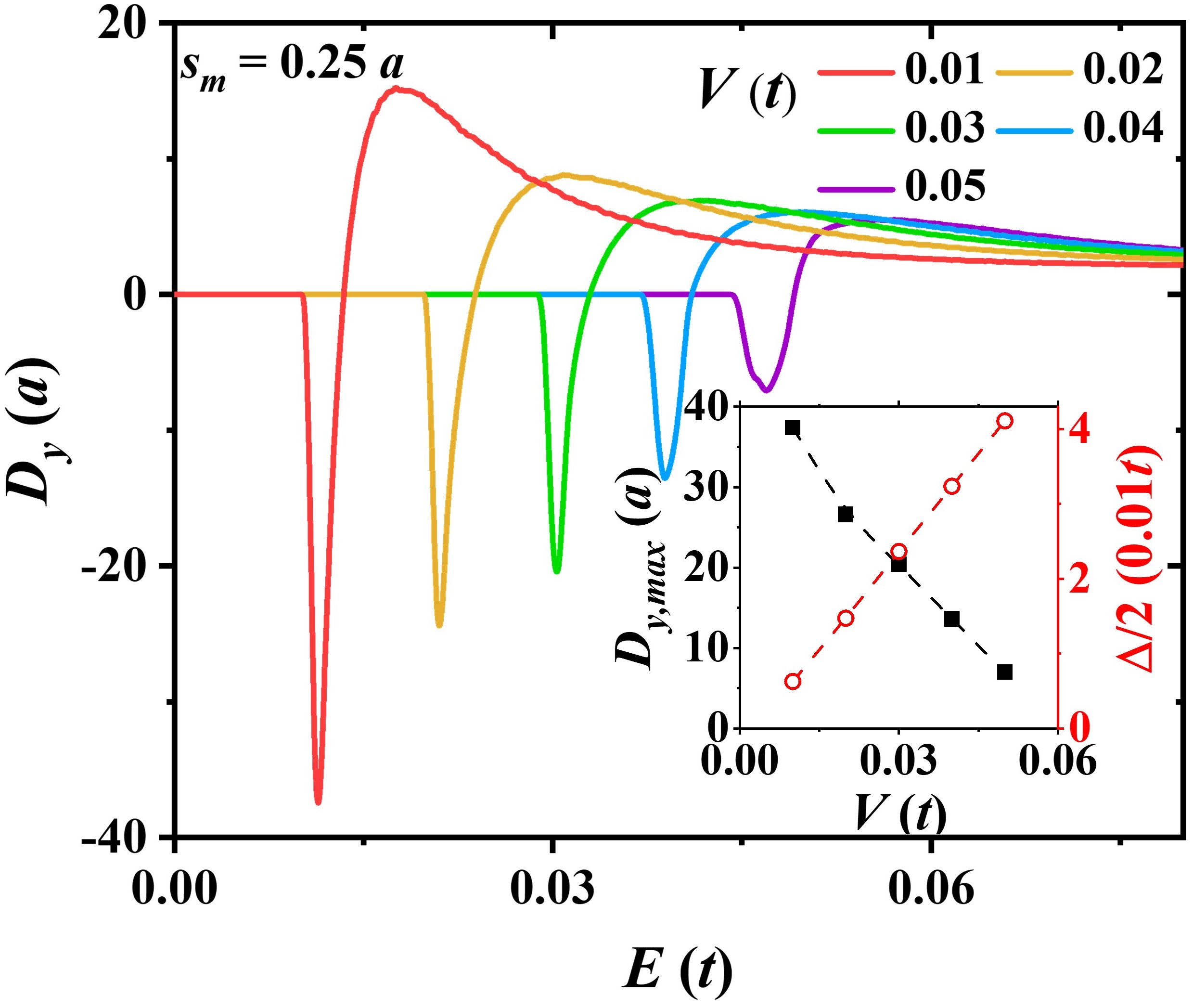}
\caption{\textbf{Simulated Berry curvature dipole under different potential differences.} Berry curvature dipole plotted as a function of Fermi energy for different potential differences with a fixed sliding distance $s_m = 0.25a$. The inset summarizes the maximum dipole magnitude $D_{y,{\rm max}}$ (black solid squares) and half of the bandstructure gap $\Delta/2$ (red open circles) as functions of potential difference. Dashed curves serve as a guide to the eye.}
\label{potential}
\end{figure}

An alternative way to manipulating the Berry curvature dipole is by varying interlayer potential difference, which can be experimentally achieved using dual gates. In this subsection, we fix the maximum interlayer sliding distance at  $s_m =0.25a$. The Berry curvature dipole for different potential differences, ranging from $0.01t$ to $0.05t$ is shown in Fig. \ref{potential}. As evident from the figure, the absolute magnitude of the Berry curvature dipole increases monotonically as the potential difference decreases. To quantify this behavior, we use the maximum absolute magnitude $D_{y,{\rm max}}$ to characterize the overall trend. The values of $D_{y,{\rm max}}$ for different potential differences are summarized in the inset in Fig. \ref{potential}, shown as solid black squares. The increase in $D_{y,{\rm max}}$ is attributed to the fact that a larger Berry curvature is confined within a more compact $k$ space for weaker potential differences, similar to the observations reported in previous studies\cite{Battilomo2019,pan2024}.

It is important to note that while the theoretical maximum value of the Berry curvature dipole appears at very weak interlayer potential difference, achieving a large Berry curvature dipole experimentally is challenging due to smearing effects from temperature and substrate disorder. The latter is primarily influenced by the typical energy scale induced by disorder in substrate; for instance in SiO$_2$, it is around $\sim 0.01t$ \cite{Martin2008}, comparable to the energy scale considered in this study. The relevant energy scale against smearing effect of disorder, can be modeled by half of the band gap, $\Delta/2$. We summarize $\Delta/2$ as a function of potential difference in the inset of Fig. \ref{potential}, as shown by the open red circles \bibnote{The half gap $\Delta/2$ does not coincide with the onset energy of the Berry curvature dipole due to the absence of particle-hole symmetry in this system.}. Consequently, although the Berry curvature dipole is larger at weaker potential differences, the dipole generated at larger potential differences is more robust against the smearing effects induced by disorder of the corrugated substrate.

\subsection{\comment{Berry curvature dipole in trilayer and tetralayer graphene}}

\begin{figure} [t]
\includegraphics[width =1\linewidth]{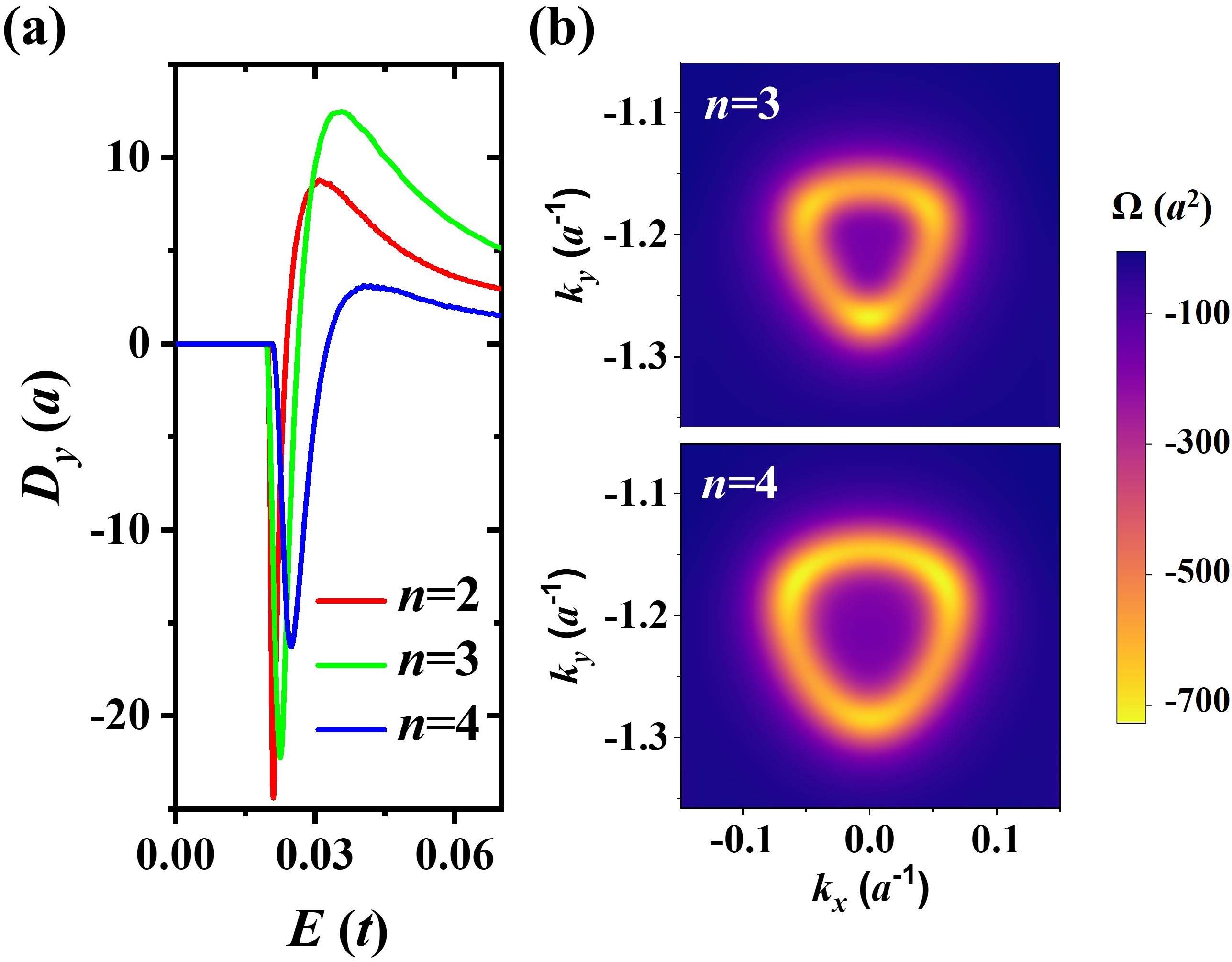}
\caption{\comment{\textbf{Berry curvature dipole for bilayer, trilayer and tetralayer graphene.} (a) Berry curvature dipole $D_y$ as a function of Fermi energy for different layer numbers $n=2, 3, 4$. The layer potential difference $V$ is set to $0.02t$, $0.01t$ and $0.007t$ for bilayer, trilayer and tetralayer graphene respectively, a consistent maximum sliding distance of $s_m=0.25a$. (b) Berry curvature for trilayer and tetralayer graphene are shown in top and bottom panels.}}
\label{layer}
\end{figure}

\comment{In this subsection, we extend our analysis from bilayer graphene to trilayer and tetralayer graphene. Similar to the bilayer case, it is necessary to open a band gap near the Dirac point for multilayer graphene. For trilayer graphene without interlayer sliding, there are two common stacking configurations: Bernal (ABA) and rhombohedral (ABC). Although the ABA stacking is more commonly observed, it does not permit the opening of a band gap under a perpendicular displacement field \cite{trilayer1,trilayer2,multilayer1,multilayer2,multilayer3}. Therefore, we focus on the ABC stacking, where a band gap can be effectively induced by applying displacement fields \cite{trilayer1,trilayer2}. For tetralayer graphene, three stacking configurations exist: ABAB, ABAC (or equivalently ABCB), and ABCA. Among these, we focus on the rhombohedral ABCA stacking due to its ability to open a sizable band gap with displacement field\cite{multilayer1,multilayer2,multilayer3}.  To ensure a comparable analysis, we assign $V=0.02t, 0.01t, 0.007t$ for bilayer, trilayer and tetralayer graphene respectively, such that these configurations exhibit Berry curvature dipoles with similar onset energies as shown in Fig. \ref{layer}(a).}

\comment{The appearance of a nonzero Berry curvature dipole in both trilayer and tetralayer graphene arises from the breaking of three-fold rotational symmetry, similar to the bilayer graphene case. We have confirmed that for multilayer graphene without interlayer sliding, the Berry curvature retains three-fold rotational symmetry, resulting in a zero Berry curvature dipole. When inhomogeneous sliding is introduced, the Berry curvature density no longer exhibits three-fold rotational symmetry, as illustrated in Fig. \ref{layer}(b). For the trilayer graphene case (top panel), the lower Berry curvature "hotspot" has a larger magnitude than the upper two. Conversely, in the tetralayer graphene case (bottom panel), the opposite pattern is observed. In addition to the breaking of three-fold rotational symmetry, the Berry curvature density retains mirror symmetry with respect to $k_x=0$, owing to the preserve of time-reversal symmetry in $x$ direction. Consequently, the $x$ component of Berry curvature dipole $D_x$ is zero for multilayer graphene, leaving only the $y-$component $D_y$ nonzero.}

\comment{Though the appearance of a nonzero Berry curvature dipole underscores the generality of our findings across multilayer graphene systems, it is important to note that as the layer number $n$ increases, the bandstructure dispersion follows $k^n$, resulting in stronger correlation effects. In rhombohedral multilayer graphene, strong correlation effects have been observed\cite{rhombohedral1}, giving rise to phenomena such as superconductivity \cite{rhombohedral-sc1,rhombohedral-sc2}, multiferroicity\cite{rhombohedral-multiferroicity} and, more recently, the fractional quantum anomalous Hall effect\cite{rhombohedral-fqahe}. These experimental findings suggest that a single-particle picture is insufficient to fully capture the underlying physics of rhombohedral multilayer graphene. How inhomogeneous interlayer sliding influences these correlation effects remains an open question, warranting further experimental and theoretical investigation in the future. }

\subsection{Size scaling of Berry curvature dipole}

\begin{figure} [t]
\includegraphics[width =1\linewidth]{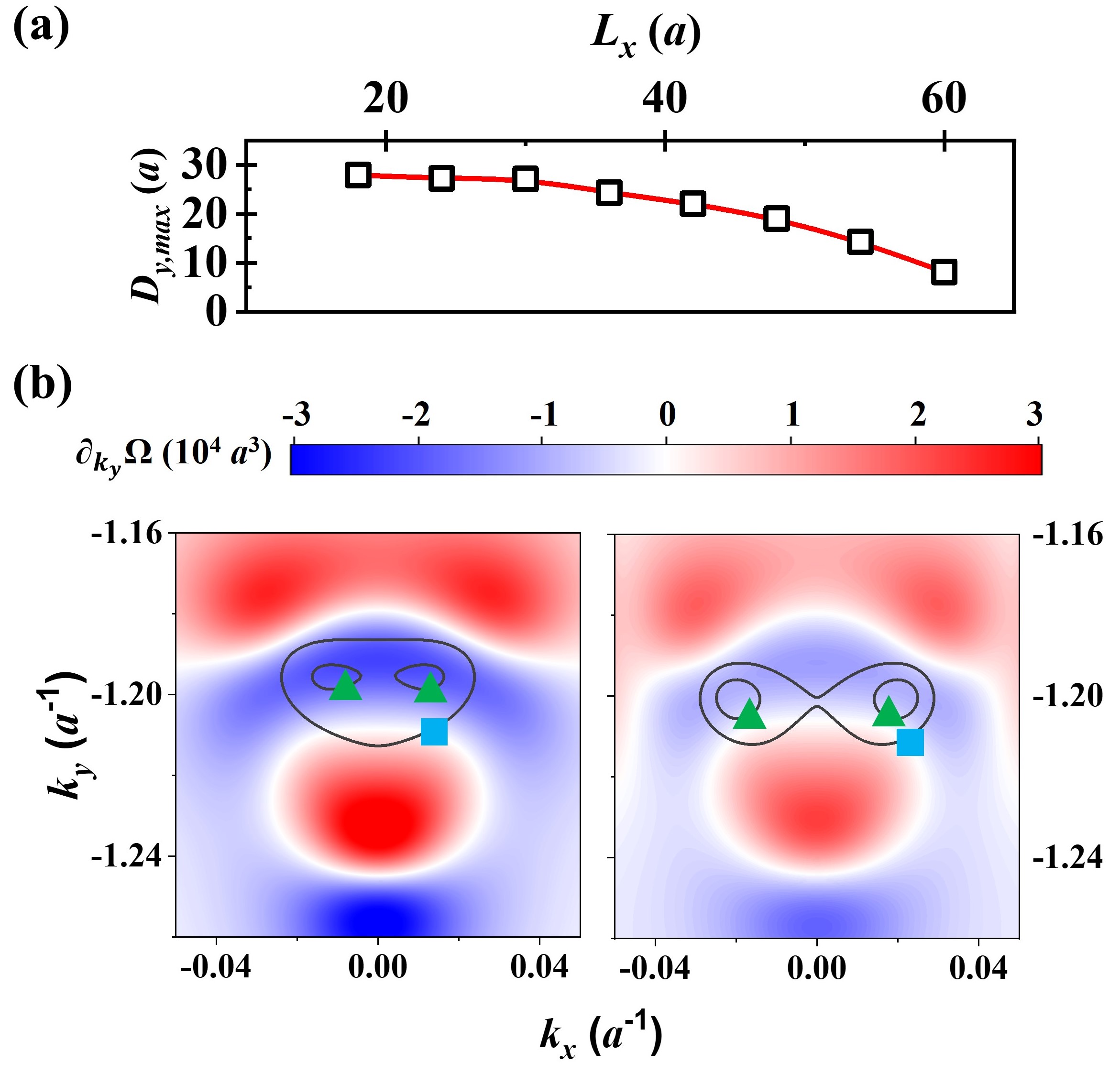}
\caption{\textbf{Size scaling behavior of Berry curvature dipole.} (a) Dependence of the Berry curvature dipole on length scale $L_x$. Symbols represent simulated results, with the red curve serving as a guide to the eye. (b) Berry curvature densities for $L_x =24a$ and $L_x=54a$ are shown in left and right panels, respectively. Triangular symbol denotes the iso-energetic lines of near the bottom of conduction band, while square symbol mark the critical iso-energetic lines where the maximum absolute magnitude of the Berry curvature dipole is reached.}
\label{length}
\end{figure}

In this subsection we examine the size scaling behavior of Berry curvature dipole in bilayer graphene. We use the maximum magnitude to characterize Berry curvature dipole variation. Here, the interlayer sliding distance $s_m$ is fixed at $0.25a$ and potential difference is set to $V=0.02t$. The periodicity $L_x$ , defined as the length scale in Fig. \ref{cartoon}(b), plays a crucial role. A clear decline $\sim 70\%$ of Berry curvature dipole magnitude is observed as the periodicity $L_x$ increases from $18a$ to $60a$, as illustrated in Fig. \ref{length}(a). 

We attribute the decrease of Berry curvature dipole to two effects. (i) The Berry curvature dipole density decreases with the periodicity. This point can be seen by comparing left ($L_x=24a$) and right ($L_x=54a$) panels in Fig. \ref{length}(b), where the left one's Berry curvature dipole density has smaller absolute magnitude. Our simulation results show a 30\% decrease of dipole density as the periodicity increases from $18a$ to $60a$. (ii) Bandstructure variation. Changes in the bandstructure with increasing periodicity, also contribute to the decline. To illustrate this effect, we compare cases with $L_x=24a$ and $L_x=54a$ in Fig. \ref{length}(b). In both cases, the Berry curvature dipole density and bandstructure exhibit the mirror symmetry with respect to $k_x=0$. For shorter periodicity $L_x=24a$, the bottom of conduction band (labeled by the triangle) appears closer to center region (where the magnitude of Berry curvature dipole density is larger), as compared with $L_x=54a$. Similar behavior is seen for critical energy labeled by the square ($D_{y,{\rm max}}$ appears at this energy point), the shorter periodicity case occupies region closer to center. This contributes to an additional decline in the Berry curvature dipole as the periodicity increases.

The typical length scale of the Berry curvature pocket in $k$ space of our bilayer graphene system is around $0.1a^{-1}$, as illustrated in Fig. \ref{bc}. This corresponds to a real-space length of approximately $20\pi a\sim 60a$. Consequently, when the length scale increases beyond $L_x>60a$, we must consider the second conduction band, which is found to intersect with the third one. This intersection leads to a divergence in the Berry curvature, as defined by Eq. \ref{equ_bc}. The resulting band degeneracy suggests that the system enters a non-Abelian regime \cite{nonAbelian,nonAbelian2}, which requires further theoretical investigations for systems with larger periodicity.

\section*{IV.	Summary}

In summary, we proposed an experimentally feasible method for generating inhomogeneous interlayer sliding in \comment{multilayer} graphene by placing it on a corrugated substrate. Our findings demonstrate that this inhomogeneous sliding effectively breaks the three-fold rotational symmetry, resulting in nonzero Berry curvature dipoles. Furthermore, when interlayer potential differences are applied, a nonzero Berry curvature dipole is generated. Our simulations show that the Berry curvature dipole can be effectively tuned by varying the interlayer sliding distances and potential differences. Notably, the maximum inhomogeneous sliding distance, $s_m$, can lead to a Berry curvature dipole magnitude as large as $D_y \sim 100 s_m$, highlighting \comment{inhomogeneous} interlayer sliding as an \comment{experimentally realizable} method for generating substantial Berry curvature dipoles.

\section*{Acknowledgment}
This work is financially supported by National Natural Science Foundation of China (Grants no. 12304232 and 12374121), Shaanxi Fundamental Science Research Project for Mathematics and Physics (Grants no. 22JSY026, 23JSQ011), the Fundamental Research Funds for the Central Universities.
\bibliography{biblio.bib}

\end{document}